# About the waiting time for a strong earthquake

A.V. Guglielmi, and O.D. Zotov

*Institute of Physics of the Earth, RAS, Moscow, Russia*

*guglielmi@mail.ru, ozotov@inbox.ru*

**Abstract**

As an object of study, we chose the global activity of strong earthquakes (M > 7). The subject of the study is the waiting time for the next strong earthquake. The purpose of the study is to compare two distributions of waiting time, one of which is calculated when ordering events according to world time, and the second when ordering according to proper time. We have come up with unique underground "clock" to counting proper time. The average waiting time is 25 days in world time, and 10 ticks in proper time. We assert with a high degree of confidence that both distributions obey an exponential law. The coefficients of determination are 0.94 in the first case, and 0.98 in the second case. This is all the more surprising since our underground clock is a rather crude measuring instrument. We also consider general issues related to the chronologization of earthquakes.

*Keywords*: proper time, deactivation factor, main shock, foreshocks and aftershocks, exponential and power distributions.

## 1. Introduction

111 years have passed (a kind of anniversary) since the discovery of the dependence of the flow of time on the gravitational potential [1]. It was impossible not to mention this, since we intend to discuss the question of the flow of time in the physics of earthquakes. Of course, there can be no question of a deep analogy between the fundamental laws of the theory of relativity and the semi-empirical regularities that have been established so far in seismology when studying the time series of earthquakes. Nonetheless, it should be said directly that the term "proper time", which we will use in this paper, is borrowed from Einstein's theory of relativity.

The concept of the proper time of an earthquake source originally arose in the theory of aftershocks. Let us represent the Omori law [2] in differential form [3]

$$\frac{dn}{d\tau} + n^2 = 0. \qquad (1)$$

Here $n$ is the frequency of aftershocks.
The relationship between proper time $\tau$ and universal time $t$ is given by the relation $d\tau = \sigma dt$, where $\sigma$ is the deactivation factor of an earthquake source.

The idea of proper time makes it possible to describe the evolution of aftershocks using only one phenomenological parameter characterizing the source state. This is a clear advantage over the Hirano-Utsu formula [4, 5], which contains three parameters.

The evolution equation (1) has been successfully used in a number of studies (see, for example, [6–15]). In particular, it was predicted and experimentally found that the deactivation factor decreases with increasing magnitude of the mainshock [12].

Seismology, geodynamics, and tectonophysics have not yet developed a definite attitude to the question of proper time. This can be understood, since the concept of proper time in geophysics, strictly speaking, is not necessary, but we use it to briefly express an idea that requires special explanation.

It is useful to give a simple, but by no means simplistic, example. Imagine a laboratory in which the oscillations of a pendulum on a long suspension are studied. If you count time using radio signals of exact time, you will notice that the period of oscillation of the pendulum slowly changes with time. Why? Going through the possible answers, the experimenter fully admits the idea that the frequency of radio signals, generally speaking, may not have a direct relationship to the rate of time in his own laboratory. He invents his own clock, the "ticking" of which is determined by the oscillations of a pendulum on a short suspension. And he notes with satisfaction that the period of oscillation of the object of study, measured in units of proper time, does not change over time. Needless to say, in this illustrative example, we neglect damping, that the experimenter does not know the laws of mechanics, and he has no information about the changing temperature in the laboratory, which determines the length of the suspension and, accordingly, changes the oscillation period of the object of study. In this example, the non-stationarity of the medium, which affects the oscillations of the object, is implicitly taken into account by the uneven flow of proper time.

Remotely, all this resembles the state of affairs that has developed in the physics of earthquakes. We do not fully know the mechanics of earthquakes, and the non-stationarity of the geological environment, which affects the chronological sequence of the main shocks and their consequences - aftershocks, is hidden from us. Under these conditions, it may turn out that the ordering of events according to proper time will, in a certain sense, be preferable to ordering according to world time clocks. It is quite clear that only observations can answer the question of how useful such a possibility will be in practice. We will consider the planetary activity of strong earthquakes and show that the chronologization of events according to proper time is quite effective.

## 2. Waiting time according to world clock

As the research object, we will choose the global activity of strong earthquakes from 1973 to 2019, information about which is contained in the USGS/NEIC world catalog of earthquakes (https://earthquake.usgs.gov). We will call strong earthquakes, the magnitude of which is $M \geq 7$. In total, 646 strong earthquakes occurred during the specified period of time. As an initial hypothesis, we formulate the assumption that the flow of strong earthquakes is a Poisson flow. The subject of the study will be the waiting time for the next strong earthquake. The purpose of the study is to compare two distributions of waiting time, one of which is calculated when ordering events according to world time, and the second when ordering according to proper time.

So, we will present the sequence of earthquakes as a chain of instantaneous events separated by random intervals of time. The Poisson distribution has the form

$$p_k(\lambda) = \frac{\lambda^k}{k!}\exp(-\lambda). \qquad (2)$$

At $\lambda = at$, the $p_k(t)$ value is the probability that $k$ events occur during time $t$, $k = 0,1,2,3,...$ We are interested in the case $k = 0$:

$$p_0(t) = \exp(-at). \qquad (3)$$

Based on our hypothesis, we have obtained the exponential distribution of the waiting time between two strong earthquakes. The average waiting time is $\langle t \rangle = 1/a$.

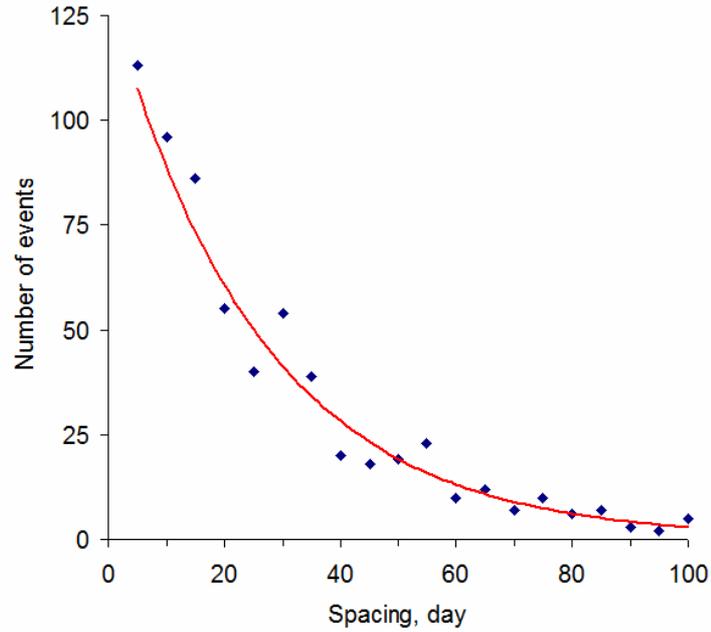

Fig. 1. Dependence of the number of events on the interval between them (black dots), and the exponential function that best approximates the experimental points (red line).

In Figure 1, the vertical axis shows the number $N$ of intervals between earthquakes, and the horizontal axis shows the waiting time $t$ for the next earthquake.

We see that the experimental points are extremely well approximated by the exponential function

$$N(t) = 139 \exp(-0.04t). \qquad (4)$$

The coefficient of determination $R^2 = 0.94$ is very large. Our hypothesis has been confirmed. The average wait time is $\langle t \rangle = 25$ days.

### 3. Waiting time according to the underground "clock"

To count the proper time, we need an "underground clock". Their course, to some extent, should reflect the state of the lithosphere, which, generally speaking, changes with the course of world time. One can imagine choices, but we are constrained in this study of the object. Namely, for completeness of the picture, it is necessary that the clock runs properly for the entire period of time indicated above. We have chosen a "clock", the ticking of which is noted by tremors of relatively low intensity. As a trial step, we chose earthquakes in the magnitude range $6 \leq M < 7$. From 1973 to 2019, 5886 such earthquakes were recorded. Information about them is contained in the USGS/NEIC catalog.

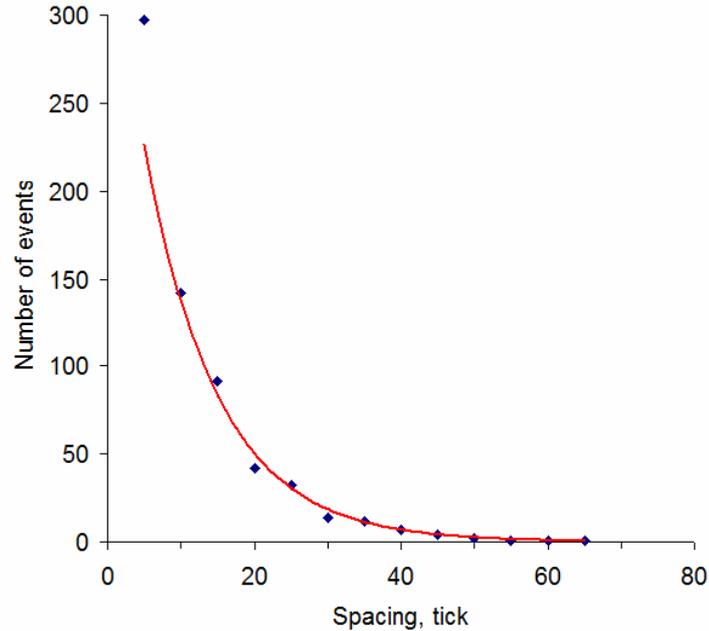

Fig. 2. Dependence of the number of time intervals between two earthquakes with magnitudes $M \geq 7$ on the interval value (black dots), and the exponential function that best approximates the experimental points (red line). In contrast to Figure 1, not world time, but proper time is used here.

The result of analysis of the waiting time for a strong earthquake in units of proper time (i.e., according to underground clocks) is shown in Figure 2. The result turned out to be amazing. We see that the exponential function

$$N(\tau) = 373\exp(-0.1\tau) \qquad (5)$$

perfectly approximates the experimental points. The average wait time is $\langle t \rangle = 10$ ticks. The coefficient of determination $R^2 = 0.98$ is even higher than when chronologizing strong earthquakes by world time clocks. It is visually obvious that the deviation of the experimental points from the exponent in Figure 2 is much less than in Figure 1. This is all the more surprising since our underground clock is a rather crude instrument.

**4. About the waiting time for foreshocks and aftershocks**

Is our stream of strong earthquakes homogeneous? Our answer is yes. When selecting earthquakes on the basis of $M \geq 7$, the database contains mainly the main shocks. Indeed, according to Bath's law [16], an admixture of aftershocks can appear only in connection with the main shocks, the magnitude of which is $M > 8$. The number of such powerful earthquakes is small with the chosen selection method. An analogue of Bath's law is known for foreshocks [11]. It imposes an even more severe restriction on the possible impurity that violates the homogeneity of the database.

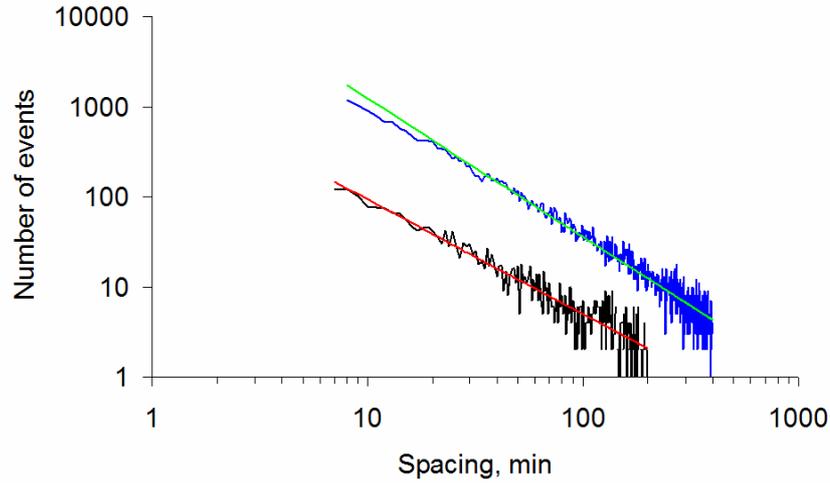

Fig. 3. Dependence of the number of events on the waiting time for foreshocks (black and red) and for aftershocks (blue and green) at mainshocks with M ≥ 6 magnitudes.

In this regard, a second question arises. Will the wait times for foreshocks and aftershocks follow an exponential law, as is the case for mainshocks? The answer turned out to be negative, as shown in Figure 3. Let us explain how this figure was obtained.

Using a simplified method [11], the main shocks with M ≥ 6 were selected. The decrease in the lower limit from M > 7 to M ≥ 6 is due to the fact that at M > 7 the total number of foreshocks is not large enough for a statistical study. We selected foreshocks and aftershocks that occurred at a depth of up to 250 km in the epicentral zones of the main shocks 24 hours before and 24 hours after the main shock. The total number of foreshocks is 3498, main shocks are 5529, and aftershocks are 35358.

Figure 3 indicates that the experimental data are well approximated by the power functions

$$N(\tau) = 1742/t^{1.27} \qquad (6)$$

for foreshocks and

$$N(\tau) = 42251/t^{1.54} \qquad (7)$$

for aftershocks. The coefficients of determination are $R^2 = 0.84$ for foreshocks and $R^2 = 0.94$ for aftershocks.

### 5. Discussion

We will not appeal to the well-known feeling that time sometimes "stretches painfully long", and sometimes "flies by rapidly". This psychological phenomenon is difficult to formalize. We have another, rather simple and clear example from classical mechanics [17]. Let us assume that a dynamic system is characterized by parameters that are determined by external fields acting on the system. If for some reason the parameters change with time, then our system is not closed, its energy is not conserved. The laws of mechanics make it possible to analyze the behavior of a system and, in particular, to find adiabatic invariants in the case when the parameters change rather slowly with time.

The situation is different in the physics of earthquakes. First, we do not know the mechanics of earthquakes. Secondly, it is extremely difficult in practice to control the time dependence of the parameters of the geological environment. These two circumstances justify our appeal to the concept of proper time in order to implicitly try to take into account the influence of the nonstationarity of the medium on the functioning of the dynamical system.

When analyzing the planetary activity of strong earthquakes, we sought and, apparently, found the only way that is currently possible for an economical and productive description of time series. Speaking about the present time, we mean that we still do not know the mechanics of earthquakes, and at the same time we must take into account the latent non-stationarity of the lithosphere. Productivity in this case simply means that our methodological setting leads to conclusions that can either be confirmed or refuted by experiment.

We also emphasized the economy of our methodology. This property is clearly manifested in our approach to ordering aftershocks in time. We base the theory of aftershocks on the Omori hyperbolic law in differential form (1). The law contains only one phenomenological parameter describing the state of the geological environment. The proper time is expressed in terms of world time using the deactivation coefficient $\sigma$, which can be relatively easily measured experimentally.

The non-stationarity of the medium is expressed in the fact that, in the general case, $\sigma$ depends on time.

It should be said with all certainty that the introduction of phenomenological parameters into the theory does not make sense if these parameters cannot be calculated on the basis of microscopic theory and/or measured experimentally. We characterize the earthquake source after the main shock by one phenomenological parameter. Without being able to make a calculation, we can nevertheless measure this parameter in each specific case. Let us change the variable $g(t) = 1/n(t)$ and rewrite the evolution equation in the form of an integral relation

$$g(t) = n_0^{-1} + \tau = \int_0^t \sigma(t') dt'. \qquad (8)$$

Differentiating (8), we get $\sigma = dg/dt$. We formally solved the inverse problem of the earthquake source in an elementary way: the value of the deactivation coefficient is obtained from observational data. To ensure the stability of the solution, it is necessary to perform regularization, which in this case is reduced to the replacement of $g \to \langle g \rangle$, where the angle brackets denote the smoothing of the $g(t)$ function. As a result, we have

$$\sigma = \frac{d}{dt}\langle g \rangle. \qquad (9)$$

Observations indicate that the deactivation coefficient changes in a complex manner with time, but at the first stage of the evolution of the usual $\sigma = \text{const}$, i.e. proper time is proportional to world time: $\tau = \sigma t$. This stage was called the *Omori epoch*. The duration of the Omori epoch varies from case to case from a few days to several months. It is during the Omori epoch that aftershocks are an exceptionally interesting object for geophysical research.

Evolution equation (1) admits interesting generalizations. For example, the logistic generalization

$$\dot{n} + \sigma n^2 = \gamma n \qquad (10)$$

contains the second phenomenological parameter $\gamma$ [18], but it, like $\sigma$, can also be calculated from observational data. The third parameter (diffusion coefficient) arose on the way of generalizing the evolution equation to the case of taking into account the spatiotemporal distribution of aftershocks [19]. This parameter is estimated from the data on the aftershock propagation velocity along the earth's surface.

## 6. Conclusion

We searched and, apparently, found two ways that make it possible to take into account the unevenness of the flow of time in the chronologization of time series of earthquakes. Our search was motivated by the general idea that the functioning of a dynamic system depends on the changing parameters of a non-stationary environment in which this system is immersed.

In earthquake physics, the idea of proper time originated during the development of the theory of aftershocks [3]. It allowed in a natural way, without violating the fundamental Omori law, to take into account the non-stationarity of the geological environment in the earthquake source, which cools down after the main shock. Developing the idea in this paper, we have shown that for the main shocks, it may be quite successful to try to introduce proper time using underground "clocks", the ticking of which is determined by relatively weaker earthquakes.

During the study, we found that the waiting time for a strong earthquake follows the exponential distribution low with exceptionally high accuracy. It can be stated with a certain degree of certainty that we are dealing with the simple Poisson flow. In other words, our sample of earthquakes is a stationary stream of homogeneous events, ordinary and without aftereffect.

Isn't the flow of main shocks a non-stationary Poisson process with an instantaneous density $\alpha(t)$? Recall the formal definition of $\alpha(t)$. Let $\mu(t)$ be the mathematical expectation of the number of events in the $(0,t)$ interval. Then $\alpha(t) = \lim[\mu(t+\delta t) - \mu(t)]/\delta t$ for $\delta t \to 0$. The $\lambda$ parameter is

$$\lambda(t) = \int_{t}^{t+\Delta t} \alpha(t') dt', \qquad (11)$$

i.e. the mathematical expectation of the number of events in the interval $(t, t+\Delta t)$ depends on the time $t$ in a non-stationary Poisson flow.

The most important property of the Poisson flow is the absence of aftereffect. This is a very non-trivial property, inherent in both simple and non-stationary flow, we intend to study in more detail in the course of further research.

*Acknowledgments*. We express our sincere gratitude to A.D. Zavyalov for valuable advice on seismology and tectonophysics. We sincerely thank B.I. Klain for discussing the physical and mathematical aspects of the work. We thank the staff of the US Geological Survey for providing earthquake catalogs.